\documentstyle[12pt,aaspp4]{aastex}

\input epsf

\begin{document}
\title{AGN Jet Mass Loading and Truncation by  Stellar Winds}
\author{A. Hubbard\altaffilmark{1,2}, E.G. Blackman\altaffilmark{1,2}}
\affil{1. Dept. of Physics and Astronomy, Univ. of Rochester,
    Rochester NY, 14627; 2. Laboratory for Laser Energetics, Univ. of Rochester, Rochester NY, 14623}

\begin{abstract}

Active Galactic Nuclei can produce extremely powerful jets.  While tightly collimated, the scale of these jets and the stellar density at galactic centers implies that there will be many jet/star interactions, which can mass-load the jet through stellar winds.  Previous work employed modest
wind mass outflow rates, but this does not apply when mass loading is provided by a small number of high mass-loss stars.
We construct a framework for jet mass-loading by stellar winds for a broader
spectrum of wind mass-loss rates than has been previously considered.
Given the observed stellar mass distributions in galactic centers,
we find that even highly efficient ($0.1$ Eddington luminosity)
jets from supermassive black holes of masses $M_{BH} \la 10^4M_{\odot}$
are rapidly mass loaded and quenched by stellar winds. For $10^4 M_{\odot}<M_{BH}<10^8 M_{\odot}$, the quenching length of highly efficient jets is independent of the jet's mechanical luminosity.
Stellar wind mass-loading is unable to quench efficient jets from
more massive engines, but can
account for the observed truncation  of the inefficient 
M87 jet, and implies a baryon dominated composition on scales $\ga 2$ kpc
therein even if the jet is initially pair plasma dominated.

\end{abstract}


\section{Introduction}

Active Galactic Nuclei (AGN) can produce extremely energetic jets.  While these jets can be very tightly collimated out to kilo-parsec scales, the volume contained by these jets is large and they will inevitably encounter many stars while passing through the dense centers of galaxies.  \citet{Komissarov94} established stellar winds as a source of jet mass-loading.  He modeled tails blown downstream but  neglected the finite size of the initial wind-blown bubble.  Komissarov found that for a total mass loading rate supplied
by  individual stars all supplying mass at the 
averaged mass-loss rate of $10^{-12} M_{\odot}/$yr, the
tails merged and spread their mass throughout the entire jet cross section.
He also found that when 
the same total mass-loading rate with a fixed average mass 
loading rate is provided by a small fraction of 
high mass-loss stars in the jet, their tails would neither merge nor cover the jet cross section.

Here however, we find that bubbles blown by stellar winds inside a jet 
can quench the jet irrespective of  the tails.  AGB stars and massive main sequence stars 
can have mass-loss rates as high as $10^{-4} M_{\odot}$/yr, and
the bubbles blown by even a small number of such stars 
can  block substantial portions of the jet \citep{compact}.  
We examine the effect of bubbles blown by a variety of stellar winds and show that while high mass, low terminal velocity winds (such as those from AGB stars)
are poorly spread  throughout the jet, such winds are nevertheless strong enough to significantly affect jet dynamics.  Bubbles efficiently spread the material of winds launched by stars with low mass-loss rates or high terminal velocities throughout the jet.

In Sec. 2  we derive  the 
conditions that determine whether stellar winds can truncate a given AGN
 jet.  These depend on the total stellar wind mass-loss rate encountered by the jet and the 
fraction of the jet cross section intercepted by each wind.  In Sec. 3 we discuss  parameter regimes of 
jet mass loading for application to specific cases.
In Sec. 4 we apply the formalism to the kilo-parsec scale jet of M87,
and   find that our estimates for the mass-loading rates required to stop the jet agree with the total amount of stellar mass-loss contained within the jet.
We conclude in Sec. 5.

\section{Conditions for jet mass truncation via stellar winds}

To study the stellar wind mass-loading of a relativistic jet we must compare the total mass-loading rate $\dot{M_T}$ of all wind sources to the jet's 
mechanical  luminosity $L_J$ (hereafter, reference to ``jet luminosity'' will
imply the total mechanical luminosity).  Taking $\gamma$ as the jet's Lorentz factor, we see that if $\gamma \dot{M_T} c^2<L_J$, the mass-loading is too weak to affect the jet while if $\gamma \dot{M_T} c^2>L_J$ the jet can be slowed or stopped.  We define
\begin{equation}
\dot{M_J}\equiv L_J/\gamma c^2
\label{Mcrit}
\end{equation}
 as the critical mass-loading rate.  If a black hole of mass $M_{BH}$ and corresponding Eddington luminosity $L_{Edd}$ launches a jet of mechanical luminosity  $0.1L_{Edd}$, $\dot{M_J}\simeq (2/\gamma) (M_{BH}/M_{\odot}) \times 10^{-10} M_{\odot}$/yr.  For $M_{BH}=10^6 M_{\odot}$ and $\gamma=2$ then, $\dot{M_J}=10^{-4} M_{\odot}$/yr.  This value of $\dot{M_J}$ is approximately the upper limit for known stellar winds, implying that such a jet could be impeded by a single object.  However, 
the condition $\dot{M_T}=\dot{M_J}$ does not suffice to slow the jet.  A second requirement is that the mass be spread throughout the jet.  We express this additional condition by considering the fractions $f$ of the jet's cross-sectional area intercepted by stellar wind bubbles. 

The
two conditions required for quenching the jet are then
\begin{equation}
\sum_{stars\ n} {\dot M}_n = {\dot M}_T \ge {\dot M}_J,
\label{mdotreq}
\end{equation}
and
\begin{equation}
\sum_{stars\  n} f_n \ge 1,
\label{mixreq0}
\end{equation}
where ${\dot M}_n$ and $f_n$ are the wind mass loss rate and wind covering fraction for the $nth$ star. 
In Eq. (\ref{mixreq0}), we have ignored  the possibility of two bubbles intercepting the same arc of the jet and the shielding of 
one stellar wind by another.

The fraction of the jet blocked by a wind will approach a maximum $f_M\equiv f(t=\infty)$ that we will show to be $f_M=\dot{M}/\dot{M_J}$ in Sec. 2.1.  This is precisely the fraction of the jet that the wind can load according to Eq. \ref{mdotreq}, and thus winds characterised by $f=f_M$ may be considered to have maximally loaded the jet.  We define an interception function for each star 
\begin{equation}
I\equiv f/f_M
\label{intercept}
\end{equation}
and use it to rewrite Eq. \ref{mixreq0} as
\begin{equation}
\sum_{stars\ n} f_n = \sum_{stars\ n} I_n f_{M, n} = {\sum_{stars\ n} I_n \dot{M_n}\over \dot{M_J}} \ge 1.
\label{mixreq}
\end{equation}
Since $I \le 1$, any system meeting the mass-mixing requirement (Eq. \ref{mixreq}) also meets the total mass-loss requirement (Eq. \ref{mdotreq}) and hence 
Eq. (\ref{mixreq}) becomes the necessary and sufficient  requirement for jet truncation
via stellar winds.

Eq. (\ref{mixreq}) allows us to relate the volume of a mass-loaded jet (through the number of stars contained therein and the stellar number density), the type of the stars (through $I$ and $\dot{M}$) and the luminosity of the jet (through $I$ and $\dot{M_J}$). 
We explicitly derive $I$ in the next section and outline an estimate for $\dot{M_T}$ in section 2.2.  In section 3.3, 
we will see that $I$ depends linearly on $\dot{M_J}$ 
for some star/jet combinations.  We can see that if such stars dominate jet mass-loading, the jet's luminosity drops out of Eq. (\ref{mixreq}).
In these circumstances, the term to the left of the inequality in 
Eq. (\ref{mixreq}) depends only on $\dot{M_T}$, as will be discussed in section 3.4.

\subsection{The interception function}

As shown schematically in Fig. \ref{diag}, the bubble blown by a wind has radius $r$ and intercepts a fraction $f=r^2/R_J^2$ of the jet.
We characterise a stellar wind by its mass-loss rate $\dot{M}$ and terminal velocity $v_{\infty}$.  The star crosses the jet perpendicularly with a velocity $v_c$.  If the jet is tightly collimated, then its radius $R_J$ will be much less than its length.  We presume that the jet's luminosity $L_J$ is evenly distributed over its cross-sectional area $A_J=\pi R_J^2$.  

As long as the stellar winds are reasonably far from the central engine,
a relativistic jet velocity will dominate the other velocities (thermal speeds, $v_{\infty}$ and $v_c$).  We consider a wind-blown bubble to be an onion-like set of concentric shells, of which only the outermost will interact with the jet.  Once a shell of mass $\Delta m$ has absorbed energy $\gamma \Delta m c^2$ from the jet, the shell will be carried away by the 
jets and thus lost from the bubble.  This allows us to ignore lateral pressure, as any portion of the stellar wind  affected by such forces would have already  been carried downstream.  Thus, the bubble will expand at its (unchanged) terminal velocity $v_{\infty}$ while simultaneously losing its outer shells to the jet.  In a time $\Delta t$, the wind absorbs an energy $\Delta E=L_J (r^2/R_J^2) \Delta t$ from the jet and accordingly loses a shell of mass $\Delta m=\Delta E/(\gamma c^2)=\dot{M_J} (r^2/R_J^2) \Delta t$.  This shell has width $\Delta r=(\Delta m/\dot{M})v_{\infty}=(r^2/f_MR_J^2) v_{\infty} \Delta t$,
where we have used $f_M={\dot M}/{\dot M_J}$ (to be justified below).  We can write the equation for the growth of the wind-bubble as follows:
\begin{equation}
\frac{dr}{dt}=v_{\infty}-{\Delta r\over \Delta t}
=v_{\infty}\left(1-\frac{r^2}{f_M R_J^2}\right).
\label{eqbasic}
\end{equation}
The above equation can be solved to give
\begin{equation}
r(t)=\sqrt{f_M} R_J \frac{e^{\eta(t)}-1}{e^{\eta(t)}+1}
\label{eq2}
\end{equation}
where
\begin{equation}
\eta(t)=2 v_{\infty} t/\sqrt{f_M}R_J
\label{eta}
\end{equation}
and $\eta^2(t)/4$ is the ratio of the jet mass flux the bubble would intercept, were it unaffected by the jet, to the stellar wind's mass loss flux.
From Eq. (\ref{eq2}) we find that
\begin{equation}
f(t)=\frac{r^2(t)}{R_J^2}=f_M \left(\frac{e^{\eta(t)}-1}{e^{\eta(t)}+1}\right)^2.
\label{eqF}
\end{equation}
As in the above equation $f(\infty)=f_M$, our definition of $f_M$ is justified, and the interception function $I$ of (\ref{intercept}) becomes
\begin{equation}
I=\left(\frac{e^{\eta(t)}-1}{e^{\eta(t)}+1}\right)^2.
\label{blockingf}
\end{equation}

\subsection{The total mass loading from stellar winds and the role of the IMF}

In the absence knowing the specific stars in a jet, we can use an initial mass function $N(m)$.  In this case Eq. (\ref{mdotreq}) becomes
\begin{equation}
\dot{M_T} = \sum_{stars \  n} \dot{M_n} \simeq \int_{m_i}^{m_f} N(m)\dot{M}(m)dm \ge \dot{M_J}
\label{IMFmdotreq}
\end{equation}
and Eq. (\ref{mixreq}) becomes
\begin{equation}
\sum_{stars \  n} f_n \simeq \int_{m_i}^{m_f} N(m)I(m)\dot{M}(m)dm \ge 1.
\label{IMFmixreq}
\end{equation}
In general, $I(m)$ will depend on the age of the stellar population in question.  In Sec. 3 we  find regimes where $I$ can be approximated.  
Eq. (\ref{IMFmdotreq}) is simpler as it does not invoke $I$.


In estimating the relevant mass loss $\dot{M_T}$ of the stars contained in a jet,
 we first note that although SN deposit a lot of energy, they are not 
a significant contributor to stellar mass-loss for typical Galactic supernovae rates of 1-2 per century. We therefore assume that the averaged total mass-loss rate of a population of stars closely approximates its averaged wind mass-loss rate.  We also assume that a star loses a fraction $a \sim 1$ of its mass over its lifetime $\tau$.  The lifetime of a star behaves as $\tau \sim \epsilon m/L$ where $m$ and $L$ are its mass and luminosity and $\epsilon\equiv \tau_{\odot}L_{\odot}/M_{\odot}$ provides appropriate scaling.  Accordingly, over their entire lifetimes, stars exhibit average mass-loss rates of $\dot{M} \sim aM/\tau= aL/\epsilon$.  
While very low mass-stars lose only a small fraction of their mass over galactic lifetimes, their aggregate luminosities and mass-losses are negligible.  

For an arbitrary IMF $N(m)$, the total stellar luminosity and total averaged mass-loss  rates from a population of stars are then
\begin{equation}
L_T = \int_{m_i}^{m_f} L(m)N(m)dm
\label{stellarlum}
\end{equation}
and
\begin{equation}
\dot{M_T}=\int_{m_i}^{m_f} a \frac {m}{\tau}N(m)dm=\int_{m_i}^{m_f} \frac{a}{\epsilon} L(m)N(m)dm=\frac {a}{\epsilon} L_T=a \frac{M_{\odot}}{\tau_{\odot}} \frac{L_T}{L_{\odot}}\simeq a \times 10^{-10} \frac {L_T}{L_{\odot}} M_{\odot}/ \rm{yr}.
\label{MTdot}
\end{equation}
Remarkably, $N(m)$ does not affect the final form of the above equation and $\dot{M_T}$ depends only on $L_T$.  Applying Eq. (\ref{MTdot}) to the Milky Way, we find a stellar mass-loss rate $\dot{M_T}\sim 4 M_{\odot}$/yr, which is comparable to estimates for the observed star formation rate (e.g. \citet{SFR}).  This justifies use of Eq. (\ref{MTdot}) for the total stellar mass-loss rate.  

Despite not affecting Eq. (\ref{MTdot}), the form of $N(m)$ can influence the 
effect that stellar wind mass-loading has on a jet through  $I$ (via $v_{\infty}$ and $\dot{M}$).  In addition, the total mass loss can be dominated by a small number of massive winds.  If this number is small enough, the process of estimating 
$\dot{M_T}$ by averaging and summing over the winds contained in the jet over long periods of time will not accurately describe $\dot{M_T}$ at any given instant.  
If the condition given by Eq. (\ref{mixreq}) is on average marginally fulfilled for the characteristic jet length, significant fluctuations of the stellar population summed over in Eq. (\ref{mixreq}) can create significant jet length fluctuations.

\section{Parameter Regimes for Analysing Wind Mass Loading and Jet Truncation}

While Eq. ($\ref{mixreq}$) provides the  fundamental condition 
for whether stellar wind mass-loading can truncate jets, its implementation is complicated by the need to accurately assess the wind parameters of the constituent stars and  individual jet/wind interactions.  Here  we examine some specific ranges that parameters in Eq. (\ref{mixreq}) can take in order to understand the effects of individual jet/wind interactions as well as to find regimes in which global effects can be easily calculated.

\subsection{Strong winds ($f_M>1$)}

For a jet containing a stellar wind with $f_M>1$,  we can use Eq. (\ref{eq2}) to calculate the time $t_b$ it will take for the associated wind bubble to grow and intercept the entire jet.
Setting $r=R_J$ in Eq. (\ref{eq2}) gives
\begin{equation}
t_b=\frac{\sqrt{f_M}R_J}{2 v_{\infty}} Ln\left(\frac{1+1/\sqrt{f_M}}{1-1/\sqrt{f_M}}\right).
\label{tblock}
\end{equation}
As $f_M$ increases in the regime $f_M>1$,  $t_b$ rapidly approaches $R_J/v_{\infty}$, and the bubble  expands essentially unimpeded.  If $v_{\infty}>v_c$, the star will eventually intercept the entire jet.  If instead $v_{\infty}<v_c$ the wind will have insufficient time to intercept the entire jet and must be treated 
as winds with $f_M<1$, which we discuss next.

\subsection{Weak winds ($f_M<1)$}

Here we  examine the bubble after a time $t=t_c=R_J/v_c$ required for the star to cross halfway through the jet.  Accordingly, Eq. (\ref{eta}) becomes
\begin{equation}
\eta_c\equiv\eta(t_c)=\frac {2}{\sqrt{f_M}} \frac{v_{\infty}}{v_c}.
\label{eqEta}
\end{equation}
Substituting this into Eq. (\ref{blockingf}) gives the characteristic $I_c$ for weak winds:
\begin{equation}
I_c=\left(\frac{e^{\eta_c}-1}{e^{\eta_c}+1}\right)^2.
\label{bchar}
\end{equation}
In Fig. \ref{efficiency} we plot $I_c$ as a function of $\eta_c$.  If $\eta_c$ is large, $f$
 approaches its maximum value $f_M$.  This will occur for weak ($f_M \ll 1$) and fast ($v_{\infty} \gg v_c$) winds.  

\subsection{High Mixing vs. Low Mixing Winds}

There is a wide range of possible wind and jet parameters
(e.g. $\dot{M}$ can vary by $10$ orders of magnitude) 
but useful conclusions can be drawn merely by dividing the wind-jet systems into ``high mixing'' ($\eta_c>1$) and ``low mixing'' ($\eta_c<1$) categories.  For ``high mixing'' winds, $I \sim 1$ and Eq. (\ref{mixreq}) simplifies to Eq. (\ref{mdotreq}).  This ``high mixing'' category includes both low mass winds ($f_M \ll 1$) and high mass-loss, fast winds ($v_{\infty} \gg v_c$) such as Wolf-Rayet stars, which can single-handedly slow an entire jet.

Low mixing winds ($\eta_c<1$), such as those from AGB stars, have high mass loss rates but low terminal velocities.  As seen in Fig. \ref{efficiency}, 
for such  winds 
\begin{equation}
I(\eta_c) \simeq \frac {\eta_c^2}{4} = \frac {1}{f_M} \left(\frac{v_{\infty}}{v_c}\right)^2.
\label{lowb0}
\end{equation}
Unless the wind-jet interaction takes place close to a galactic center, the galaxy's velocity dispersion $\sigma$ will approximate the star's crossing velocity $v_c$.  Accordingly we rewrite Eq. (\ref{lowb0}) as
\begin{equation}
I_c=\frac{1}{f_M} \left(\frac{v_{\infty}}{\sigma}\right)^2.
\label{lowb}
\end{equation}
An interesting implication of Eq. (\ref{lowb}) is that 
the fraction of the jet blocked by a low mixing wind, $f\equiv r^2/R_J^2 =I_c f_M = (v_{\infty}/\sigma)^2$, is independent of both the wind's mass-loss rate and the jet's mechanical luminosity.  
This is because the wind-blown bubble stays small enough that the second term on the right of Eq. (\ref{eqbasic}) remains negligible.

Because $\dot{M}=10^{-4} M_{\odot}$/yr is a reasonable upper limit on wind strengths for both fast and slow wind stars, Eq. (\ref{eqEta}) demonstrates that all stars are``high mixing'' for any adequately powerful ($\dot{M_J} \ga 10^{-2} M_{\odot}$/yr) jet.  However, stellar winds do not provide adequate mass-loading rates to substantially slow such jets.  Stellar winds are able to slow weaker jets, but the broad range of values that $I$ can take for  ``low mixing'' winds
that enter the jet requires careful modeling 
 to use Eq. (\ref{mixreq}) in a meaningful way. 

\subsection{Jet luminosity regimes}

As seen above, the jet strength $\dot{M_J}$ does not influence Eq. (\ref{mixreq}) for ``low mixing'' winds.  If we presume that $\sum I_n {\dot M}_n$ is dominated by ``low mixing'' winds from AGB stars, Eq. (\ref{mixreq})'s resulting lack of dependancy on $\dot{M_J}$ allows us to draw general conclusions from Eq. (\ref{mixreq}) without knowing the behaviour of $I$ in detail.  We use Eq. (\ref{lowb0}) to rewrite Eq. (\ref{mixreq}) as
\begin{equation}
\frac{\dot{M_J}}{\dot{M}_{char}} \left(\frac{v_{char}}{\sigma}\right)^2  \frac{\dot{M_T}}{g \dot{M_J}} \ge 1
\label{app1}
\end{equation}
where $\dot{M}_{char}$ and $v_{char}$ are the mass-loss rate and terminal velocity of the characteristic peak mass-loading winds.  The parameter $g$ approximates the inverse 
fraction of the total mass-loss contributed by such winds.  Galactic stellar velocity dispersions satisfy $\sigma \sim 200-300$km/s.  For winds from AGB stars, ${\dot M}_{char} \sim 10^{-6} - 10^{-4} M_{\odot}$/yr and $v_{char}\sim 10$km/s.  We therefore find that
\begin{equation}
\dot{M_T} \simeq 6 \times 10^{-3} g \left({{\sigma}\over 250{\rm km/s}}\right)^2
\left({v_{char}\over 10{\rm km/s}}\right)^{-2}
 \left({{\dot{M}_{char}\over {10^{-5}} M_{\odot}/{\rm yr}}}\right)  M_{\odot}/\rm{yr}
\label{MTl}
\end{equation}
 approximates the total mass-loss required to truncate such a jet.  This calculation applies to jets with $\dot{M_J} \la 10^2 \dot{M}_{char}$.  For significantly weaker jets, the assumption that  $\Sigma I_n \dot{M_n}$ is dominated by
a narrow range of stars
 breaks down and the jet can be more easily quenched (although the quenching depends sensitively on  $I$).  Nevertheless, 
for an intermediate range of jet luminosities,  Eq. (\ref{MTl})
does apply. In this regime, the stellar mass-loss required to truncate the jets 
is independent of the jet's luminosity 
for mass-loading dominated by ``low mixing'' winds.

We are now able to define three jet luminosity regimes, characterised by 
$\dot{M_J}$. (1) Strong jets, with $\dot{M_J} \ga 10^{-2} M_{\odot}$/yr, experience all winds as ``high mixing.'' In this case Eq. (\ref{mixreq}) can be
approximated by Eq. (\ref{mdotreq}). However, these jets are so strong
that stellar winds do not represent a viable mechanism for their truncation. 
(2) Intermediate strength jets, with
 $10^{-6} M_{\odot}$/yr $\la \dot{M_J} \la 10^{-2} M_{\odot}$/yr, may have their mass-loading dominated by ``low mixing'' stars.  If so, such jets will propagate until the total wind mass-loading rate into the jet is ${\dot M}_T\sim 10^{-2} M_{\odot}$/yr (Eq. \ref{MTl}). (3) Low luminosity jets with $\dot{M_J}<10^{-6} M_{\odot}$/yr are too weak to withstand stellar wind mass-loading for long, but the details of the loading depend more 
strongly on the stellar population than the other two categories.

\section{Application to observations}

As long as the total mass-loss of  stars beyond the jet's truncation length remains significant, stellar wind mass-loading is a good candidate for jet truncation.  The stars enter the jet from the side and would likely induce an
more extended but staccato 
truncation of the jet compared to jet truncation via ISM pile up.
Both forms of mass loading would imply a baryonic jet composition at 
the truncation scale.
Once mass loading reduces the longitudinal
velocity of the jet to  its internal sound speed, the jet will
expand into a quasi-spherical lobe. 
In this respect, radio lobes are a natural
prediction of jet truncation via mass loading.

If jet truncation via stellar winds is assumed, then 
Eq. (\ref{mixreq}) can be used to constrain  the jet properties and the stellar population it contains.  In particular, by assuming the left side
of Eq. (\ref{mixreq}) exceeds unity, 
we can use any two of $\dot{M_J}$, $\dot{M_T}$ and $I$ to calculate the third.

Alternatively, one can use Eq. (\ref{mixreq}) to 
evaluate the effect that stellar wind mass-loading has on the jet by knowing the jet's total mechanical luminosity $\dot{M_J}$, total mass-loading rate $\dot{M_T}$ and the contained stellar population profile (which sets $I$).  
In this case, the inequality is assessed, not assumed.

\subsection{Application to M87}

The elliptical galaxy M87 has a well studied jet with luminosity $L_J$ estimated to be between $10^{42}$ and $10^{43}$ erg/s (\citet{M87Lum2}, \citet{M87L}) and a $\gamma$ of about $6$ (\citet{m87lor}).  It follows that $\dot{M_J} > 3 \times 10^{-6} M_{\odot}$/yr.  This value for $\dot{M_J}$ is interesting as it can be attained and indeed surpassed by single stars and suggests that the jet is in our intermediate luminosity regime (see Sec. 3.4).  The collimated jet has a  length of $2$ kpc (\citet{knots}), and is therefore truncated 
well  within the confines of M87. Stellar wind mass-loading is therefore
a plausible cause of the jet's truncation (implying the jet could be
longer lived that its length alone would indicate). 

 \citet{faber1997} provide values for the luminosity density near the center of M87.  
The jet  contains a total stellar luminosity of roughly $2 \times 10^7 L_{\odot}$ in the V-band  
and so, by Eq. (\ref{MTdot}), a stellar mass loss of $\dot{M_T} \sim 10^{-2} M_{\odot}$/yr.  This fits our estimate of the total mass-loss rate $\dot{M_T}$ required to quench intermediate luminosity jets from Eq. (\ref{MTl}), suggesting that ``low mixing'' stars dominate the mass-loading process.

 \citet{M87L} find evidence that the jet may be initially pair plasma 
dominated  near the launch point, but this does not contradict a subsequent increase in baryon fraction due to mass-loading via stars on kpc scales. \citet{M87L} also 
estimate the mechanical luminosity of the M87 jet via its influence on
the radio lobes. This estimate is independent of the jet composition.

\section{Conclusion}

We have modelled jet mass-loading via stellar winds by focusing on the bubbles blown by these winds.  We have derived a simple condition 
 between the jet and wind parameters that determines whether mass-loading and can truncate jet propagation.  
From this relation, jet truncation via stellar winds can be used 
to constrain the jet's mechanical luminosity (and baryon fraction), the total mass-loss rate of the stars contained within the jet, or the population profile of those stars.  

We have applied our model to the kilo-parsec scale jet of M87, and find that
jet truncation via stellar winds provides a consistent interpretation
of the length of that jet. 
This would also imply that stellar winds at the center of that galaxy are dominated by high mass, low velocity winds, and that the jet is baryon dominated
at the scale of its truncation, regardless of its initial launch composition.

\acknowledgements{AH acknowledges financial support of a Horton Fellowship from the Laboratory for Laser Energetics.
EGB acknowledges support from 
NSF grants AST-0406799, AST-0406823, and NASA grant ATP04-0000-0016
(NNG05GH61G).

\clearpage
\begin{figure}
\plotone{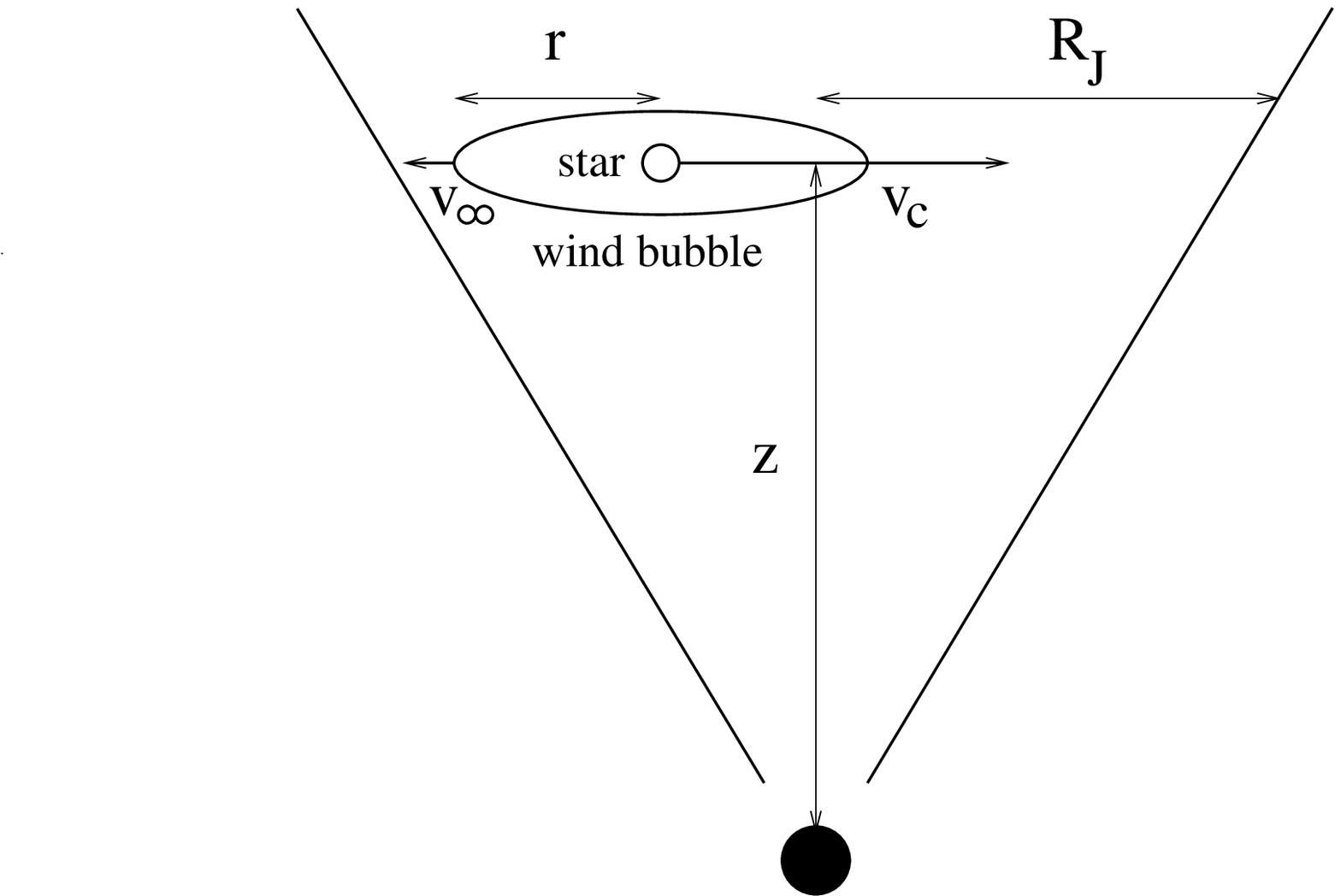}
\caption{Schematic representation of a star and wind in our model.  The star is crossing the jet, of radius $R_J$, with velocity $v_c$.  The wind, of radius $r$ is expanding at its terminal velocity $v_{\infty}$.}
\label{diag}
\end{figure}

\clearpage
\begin{figure}
\plotone{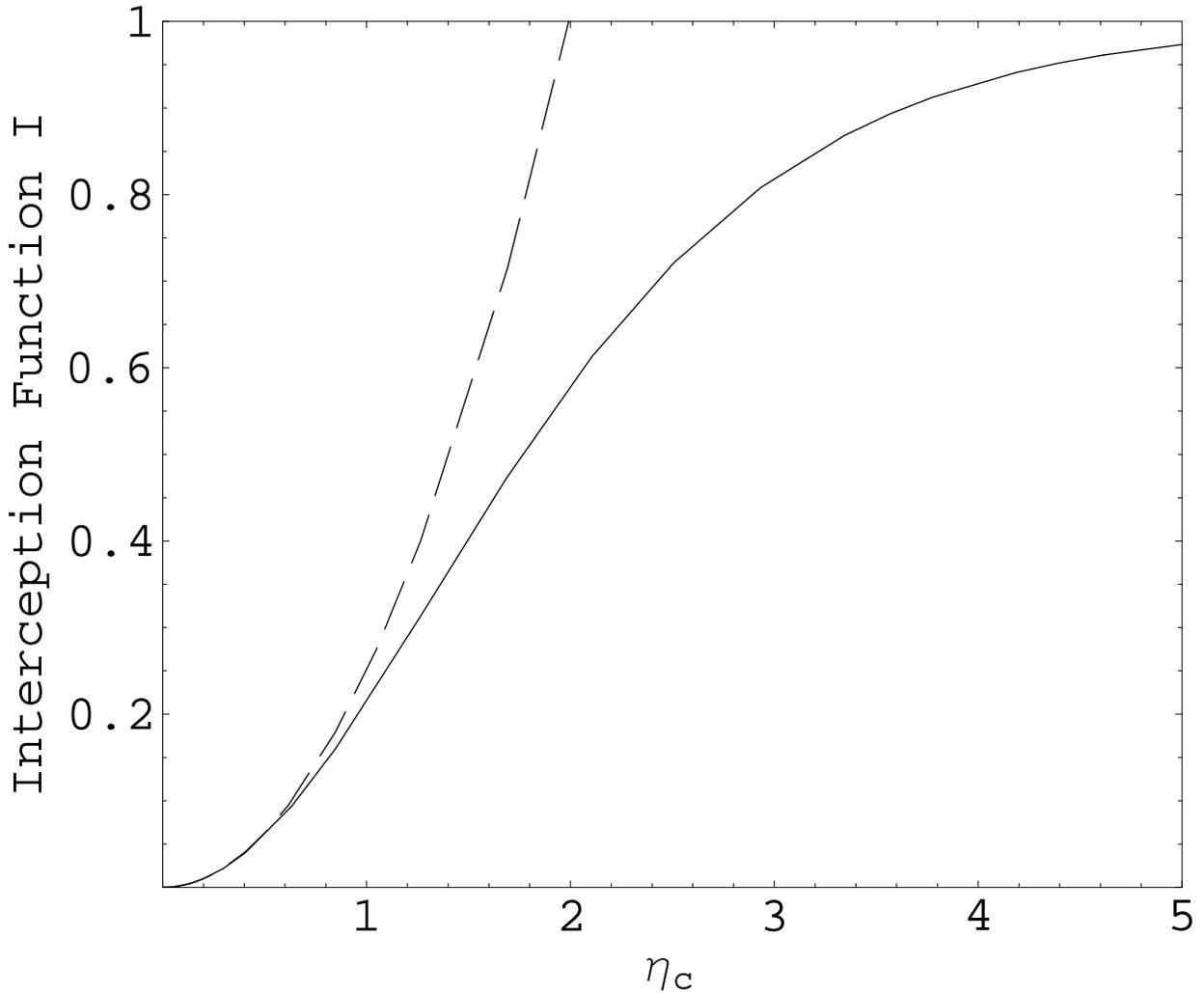}
\caption{Interception function ($I=f/f_M$) as a function of $\eta_c=2 v_{\infty}/(v_c \sqrt{f_M})$ (solid curve) as well as its low $\eta_c$ limit $\eta_c^2/4$ (dashed curve).  A value of $1$ implies that the bubble blown by the stellar wind would, having crossed halfway through the jet, have expanded to the size at which it intercepts energy flux from the jet equal to that produced by the star's mass-loss ($\gamma \dot{M} c^2$).}
\label{efficiency}
\end{figure}

\end{document}